\documentclass[a4paper]{jpconf}
\usepackage{graphicx}

\usepackage{amsfonts}
\usepackage{amssymb}
\usepackage{bm}
\usepackage{dcolumn}
\usepackage{epsfig}
\usepackage[T1]{fontenc}
\usepackage{graphicx}
\usepackage{graphics}
\usepackage{hyperref}
\usepackage[latin1]{inputenc}
\usepackage{latexsym}
\usepackage{multirow}

\newcommand{\rsu}{r^{\ast}}
\newcommand{\rsup}{r^{\ast}_{p}}

\newcommand\be{\begin{equation}}
\newcommand\ba{\begin{eqnarray}}
\newcommand\ee{\end{equation}}
\newcommand\ea{\end{eqnarray}}

\newcommand{\mb}[1]{\mbox{\boldmath $#1$}}
\newcommand{\salto}[1]{\left[\,#1\,\right]^{}_{p}}

\newcommand{\met}{\mbox{g}}

\newcommand{\singu}{{\mbox{\tiny S}}}
\newcommand{\regu}{{\mbox{\tiny R}}}
\newcommand{\Hor}{{\mbox{\tiny H}}}
\newcommand{\Inf}{{\mbox{\tiny I}}}

\newcommand{\LL}{{\mbox{\tiny L}}}
\newcommand{\RR}{{\mbox{\tiny R}}}

\begin{document}
%
\title[An Efficient Time-Domain Method to Model Extreme-Mass-Ratio Inspirals]
{An Efficient Time-Domain Method to Model Extreme-Mass-Ratio Inspirals}
\author{Priscilla Canizares$^{1}$ and Carlos F.~Sopuerta$^{2}$}
\address{Institut de Ci\`encies de l'Espai (CSIC-IEEC), \\
Facultat de Ci\`encies, Campus UAB, Torre C5 parells, 
Bellaterra, 08193 Barcelona, Spain}
\ead{$^{1}$pcm@ieec.uab.es, $^{2}$sopuerta@ieec.uab.es}

\begin{abstract}
The gravitational-wave signals emitted by Extreme-Mass-Ratio Inspirals 
will be hidden in the instrumental LISA noise and the foreground 
noise produced by galactic binaries in the LISA band. Then, we need accurate 
gravitational-wave templates to extract these signals from the noise and 
obtain the relevant physical parameters. This means that in the modeling of  
these systems we have to take into account how the orbit of the stellar-mass 
compact object is modified by the action of its own gravitational field.  
This effect can be described as the action of a local force, the {\em self-force}. 
We present a time-domain technique to compute the self-force for geodesic eccentric 
orbits around a non-rotating massive black hole.  To illustrate the method we
have applied it to a testbed model consisting of scalar charged particle orbiting a 
non-dynamical black hole.   A key feature of our method is that it does not introduce 
a small scale associated with the stellar-mass compact object.  
This is achieved by using a multidomain framework where the particle is located at 
the interface between two subdomains.  In this way, we just have to evolve 
homogeneous wave-like equations with smooth solutions that have to be communicated 
across the subdomain boundaries using appropriate junction conditions.  
The numerical technique that we use to implement this scheme is the pseudospectral 
collocation method.  We show the suitability of this technique for the modeling 
of Extreme-Mass-Ratio Inspirals and show that it can provide accurate results for the 
self-force.
\end{abstract}

\section{Motivation} \label{intro}
Thanks to the development of Gravitational Wave Astronomy, the possibility
of studying and understanding new features of the Universe is becoming a reality.
In this regard, there is an ongoing effort to learn about all the physical aspects 
of one of the main sources of Gravitational Waves (GW) for the future space-based 
Laser Interferometer Space Antenna (LISA)\cite{LISA}: Extreme-Mass-Ratio Inspirals (EMRIs).  
EMRIs are formed when a massive black hole (MBH) located at a galactic centre, with masses 
in the range $M_{\bullet}= 10^4-10^7 M_{\odot}$, captures a stellar-mass compact object (SCO), 
with masses in the range $m_{\ast} = 1-10^2 M_{\odot}$ (white dwarf, neutron star or a 
stellar black hole).  The SCO inspirals towards the MBH following a long series of 
highly eccentric orbits that shrink due to the loss of energy and angular momentum 
through the emission of GWs.  This inspiral is produced by {\em backreaction} effects related 
to the action of the SCO's gravitational field onto its own trajectory.   
In order to model EMRI systems and their GW emission in a way that we can 
produce useful GW templates for analyzing the data stream produced by LISA observations 
we need to consider the details of the gravitational backreaction responsible of the
inspiral.   This is a very challenging task for theorists.  Nevertheless, the problem
can be simplified due to the extreme mass ratios of these systems, in the range 
$\mu=m_{\ast}/M_{\bullet} \sim 10^{-7} -10^{-3}$.  This allows us to describe them in the 
framework of BH perturbation where the SCO is modeled as an accelerated point-like mass in 
the MBH background and the gravitational backreaction is pictured as the action of a local force,
the {\em self-force}.  For the purposes of our work, we can further simplify the problem by 
studying an analogous EMRI system which consist on a scalar charged point particle, $q$, orbiting
around a Schwarzschild MBH (see, e.g.~\cite{Poisson:2004lr}) that fixes the spacetime geometry. 
In this simplified model, the inspiral proceeds due to the emission of a scalar field, $\Phi$,
generated by the SCO motion and which affects the SCO trajectory through the action of a 
self-force given by the gradient of the scalar field:
\begin{equation}
 F^{\mu} = q\, (\met^{\mu\nu} + u^{\mu}u^{\nu}) \left.
\left(\nabla^{}_{\nu}\Phi\right) \right|^{}_{\gamma} \,, ~~
u^{\mu} = \frac{dz^{\mu}}{d\tau} \,, \label{particlemotion}
\end{equation}
where $\gamma$ and $u^{\mu}$ denote the SCO's trajectory and unit four-velocity respectively. 
We use this simplified model as a test bed to illustrate the techniques that we have developed 
for self-force computations.

In the case of non-rotating MBH, the spherical symmetry of the BH spacetime provides additional
simplifications of the problem.  In particular, the scalar field can be decomposed into scalar spherical 
harmonics, $\Phi^{\ell m}(t,r)$, which are decoupled between them. On the other hand, the point-like
description of the SCO produces divergences in the retarded scalar field and hence we need to 
regularize it.  In order to do so, we use the {\em mode sum} regularization scheme~\cite{Barack:2001gx}, 
which provides an analytical expression for the singular contribution of the retarded field, $\Phi^{\singu}$, 
at the particle location.  In this way, by subtracting it from the full retarded field, $\Phi$, we obtain 
a smooth and differentiable field at the particle location,
$\Phi^{\regu} = 	\Phi-\Phi^{\singu}$.  Then, the meaningful expression for the self-force is:
\begin{equation}
\textit{F}^{\;\mu} = q\, (g^{\mu \nu} + u^{\mu} u^{\nu} ) \left.
\left(\nabla^{}_{\nu}\Phi^{\regu}\right) \right|^{}_{\gamma}\,. \label{self_force}
\end{equation}
Thus,  what we need is a (numerical) technique to compute the full retarded field, $\Phi$, and to use 
the mode sum scheme to  obtain the regularized self-force.  In what follows we report on work we 
have carried out recently on the development of a new accurate and efficient technique for self-force
computations of eccentric EMRIs~\cite{Canizares:2010yx}.  This work is the extension of a previous 
one~\cite{Canizares:2008dp,Canizares:2009ay},
where these techniques were introduced for the case of circular EMRIs.


\section{Modeling EMRI{s} using a multidomain framework: The Particle-without-Particle Formulation}
\label{pwpformulation}

Our computational scheme consists in diving the computational domain in a number of subdomains
in such a way that the SCO (the particle) is always located at the interface between two subdomains.   
This has two main advantages: (i) We avoid introducing a spatial scale to resolve 
the point particle. (ii) The equations for the scalar field, which nominally have singular source
terms, become homogeneous equations at each subdomain.  These equations, assuming that appropriate
initial date is precribed, lead to smooth solutions.  This fact translates into good 
convergence properties of the numerical method used to implement this method. 

Because of these properties we call this formalism the {\em Particle-without-Particle} (PwP) 
formulation.  We evolve the individual wave-type equations at each subdomain by using time domain 
methods, which perform well for eccentric orbits.  This technique was already implemented for the
case of circular orbits and has also been presented at the previous LISA 
Symposium~\cite{Canizares:2008dp,Canizares:2009ay}.  We have recently extended the method to
make computations also in the case of eccentric orbits~\cite{Canizares:2010yx}.  In what follows
we summarize the main results of this work.

\begin{figure*}
\centering
\includegraphics[width=0.85\textwidth]{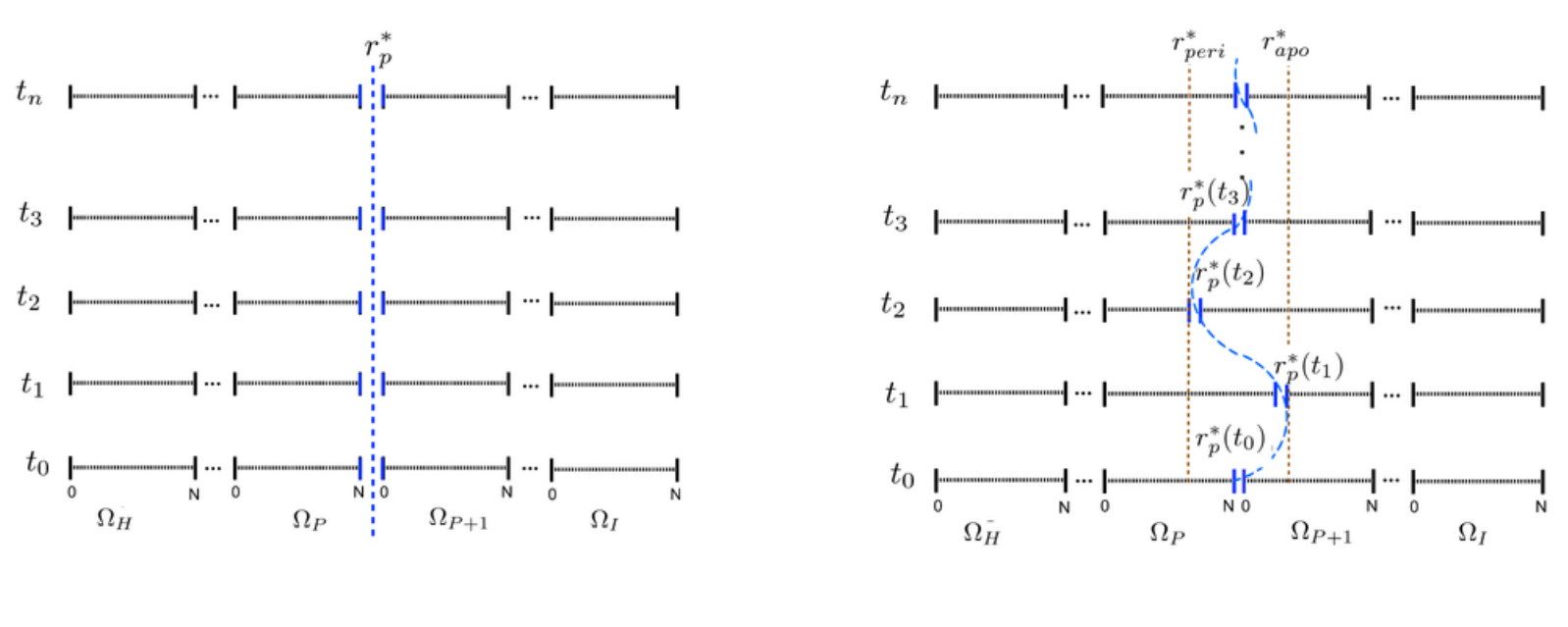}
\caption{Structure of the one-dimensional spatial domain for the circular case (left) and for 
the eccentric case (right). While in a circular orbit the particle's radial coordinate remains 
constant and the grid is static, in a generic orbit case the particle oscillates between
pericenter ($r^{}_{peri}$) and apocenter ($r^{}_{apo}$)
and the grid has to readapt continuously to maintain the particle at the interface 
between two subdomains.}
\label{multidomain}
\end{figure*}

We start by describing the multidomain structure of our PwP formulation (see Figure~\ref{multidomain}).
Once we have expanded the scalar field in spherical harmonics, each mode
satisfies an independent (not coupled to the other modes) 1+1 wave type equation
of the Regge-Wheeler type (with the potential associated to a scalar field).
Then, the spatial domain is one-dimensional: $\Omega=\left[\rsu_{\Hor},\rsu_{\Inf}\right]$, 
where $\rsu$ is the radial tortoise coordinate, $\rsu_{\Hor}$ corresponds
to the truncation in the direction to the MBH horizon and
$\rsu_{\Inf}$ corresponds to the truncation in the direction to spatial infinity.
In order to describe the subdomain communication let us consider for the moment
a splitting of the computational domain into two regions or subdomains, one to the left and one to the right
of the particle: $\rsu \in [\rsu_{\Hor},\rsu_{\Inf}] =
[\rsu_{\Hor},\rsup]\cup [\rsup,\rsu_{\Inf}]$.  Each of these regions can be in
turn divided into more subdomains  (see Figure~\ref{multidomain}):
$\Omega = \bigcup^{D}_{a=1} \Omega^{}_a$, where $\Omega^{}_a
= \left[ \rsu_{a,\LL}, \rsu_{a,\RR}\right]$ with $\rsu_{a,\RR} = \rsu_{a+1,\LL}$. 
In the eccentric case~\cite{Canizares:2010yx}, we need to use a coordinate system in which the particle
is comoving with the interface of two such domains, or in other words, the coordinates $\rsu_{a+1,\LL}$ 
and $\rsu_{a,\RR}$ associated with the particle must be time dependent.

In this setup, we use reduction 
the wave-like equation  to a first-order system by promoting the time and radial derivatives 
to new variables.  Then, these variables, $\mb{U}$, can also be split as follows: 
\begin{eqnarray}
\mb{U}(t,\rsu) &=& \mb{U}^{}_{-}(t,\rsu)\Theta(\rsup(t)-\rsu)
+\mb{U}^{}_{+}(t,\rsu)\Theta(\rsu - \rsup(t))\,,
\label{globalsolution}
\end{eqnarray}
where $\Theta$ denotes the Heaviside step function.  We can define the jump across the particle's
location in an arbitrary quantity as: $\salto{\lambda} = \mathop{\lim }\limits_{\rsu \to \rsu_{p}}\lambda^{}_{+}(t,
\rsu)- \mathop{\lim }\limits_{\rsu \to \rsu_{p}}\lambda^{}_{-}(t, \rsu)$.  We can
obtain expressions for the jumps by inserting (\ref{globalsolution}) into the field
equations (see~\cite{Sopuerta:2005gz} and~\cite{Canizares:2010yx}).  Using these junction
conditions we communicate the solutions of the homogenous equations at each subdomain.

\begin{figure}[htp]
\centering  
\includegraphics[width=0.45\textwidth]{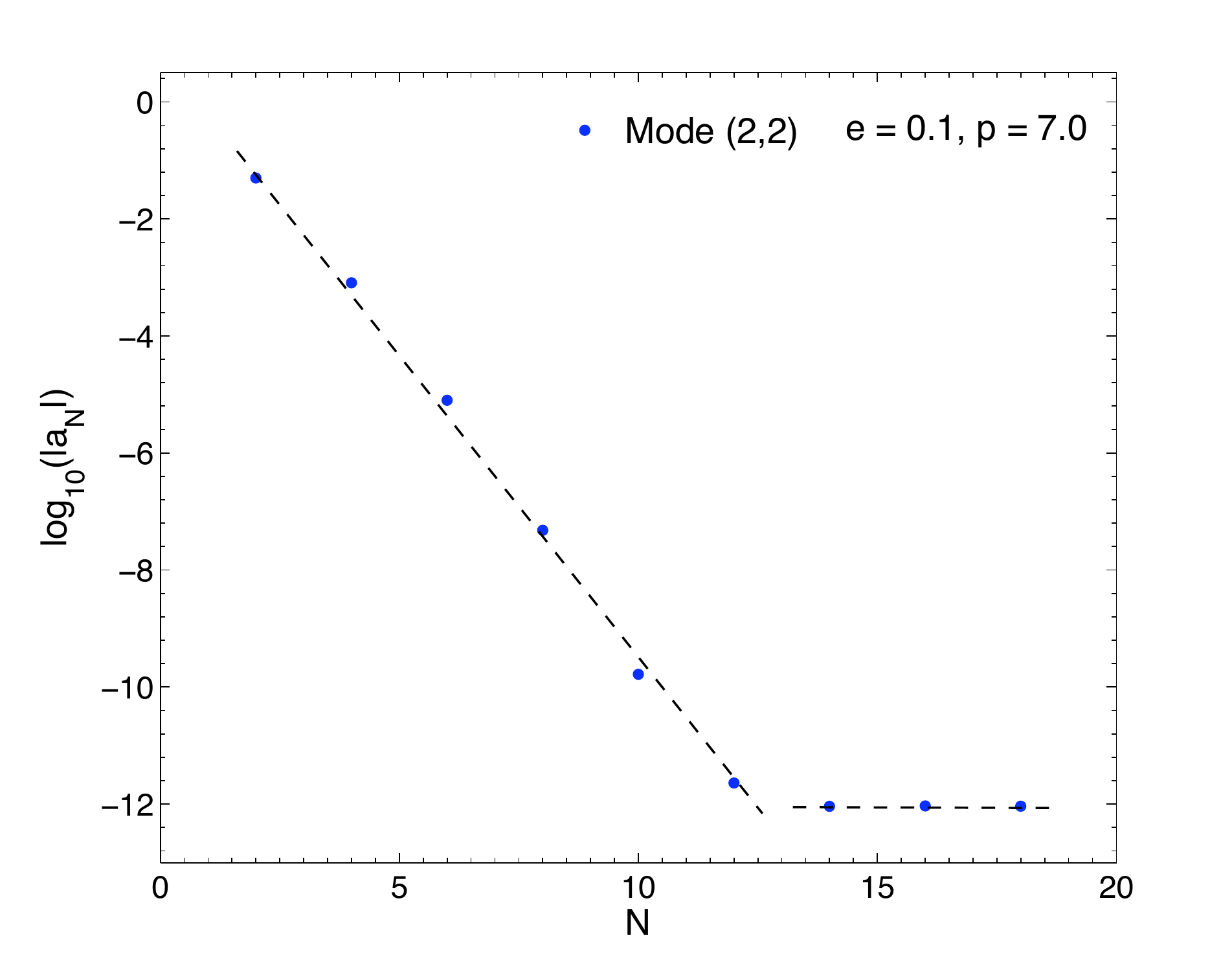}
\includegraphics[width=0.45\textwidth]{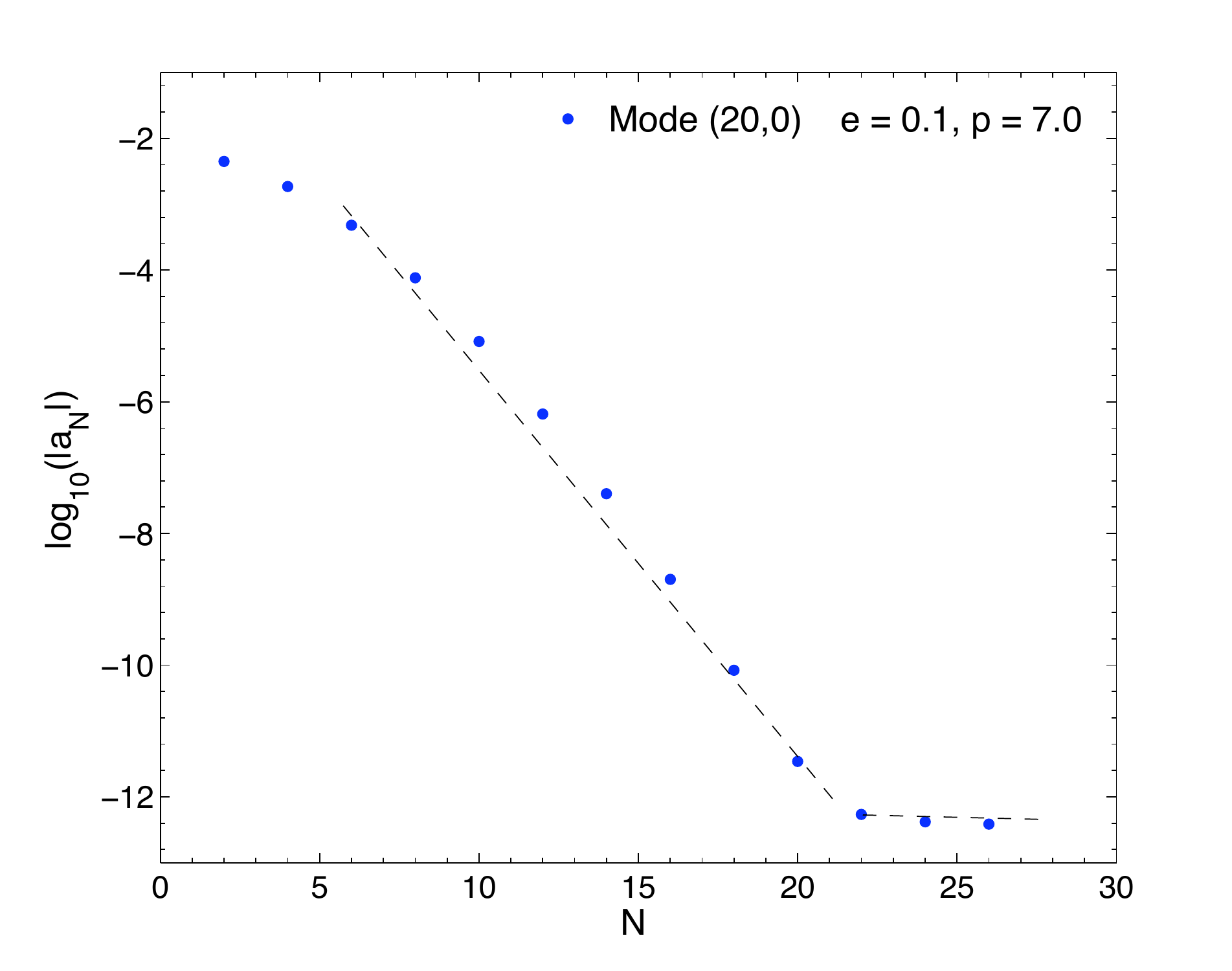}
\includegraphics[width=0.45\textwidth]{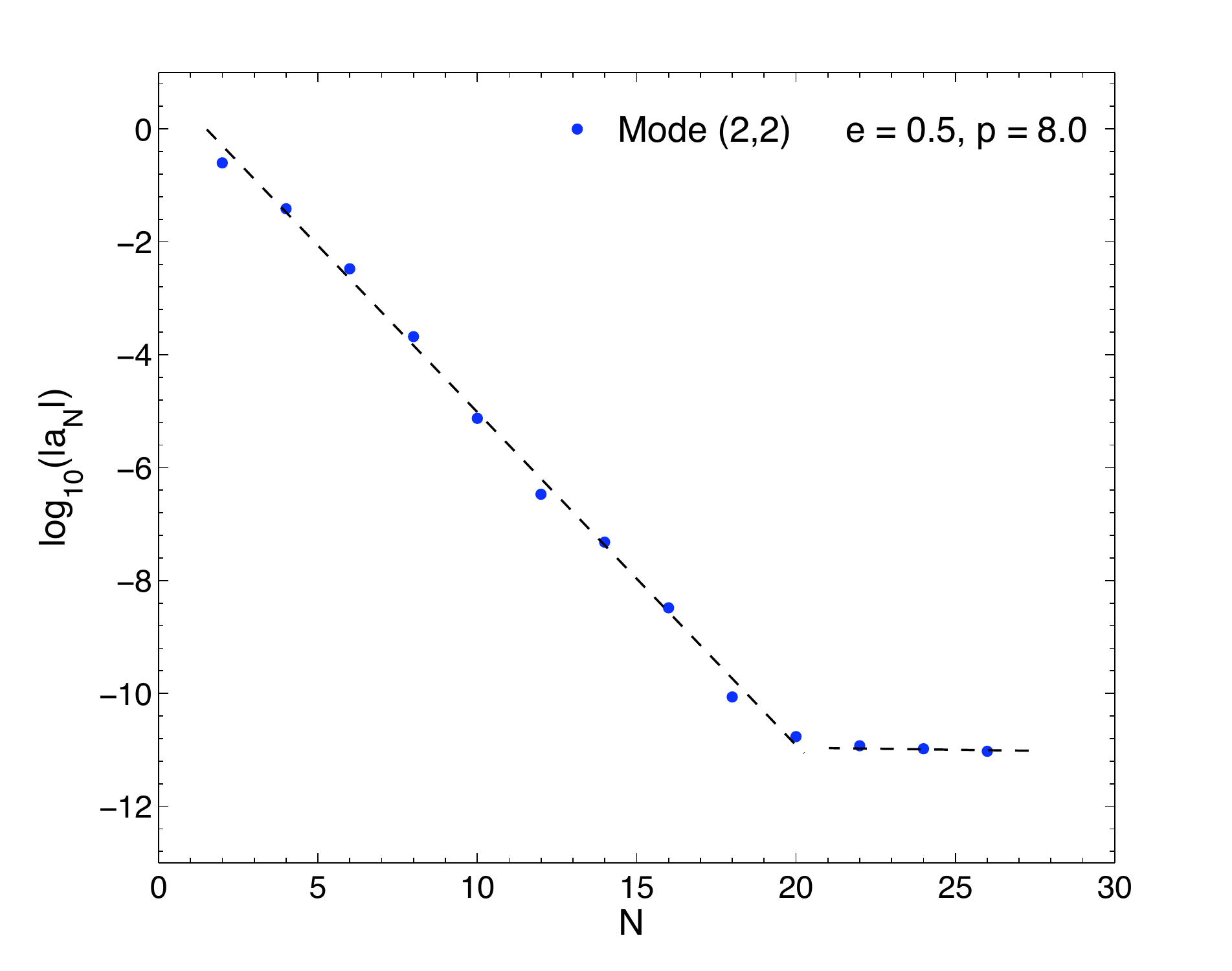}
\includegraphics[width=0.45\textwidth]{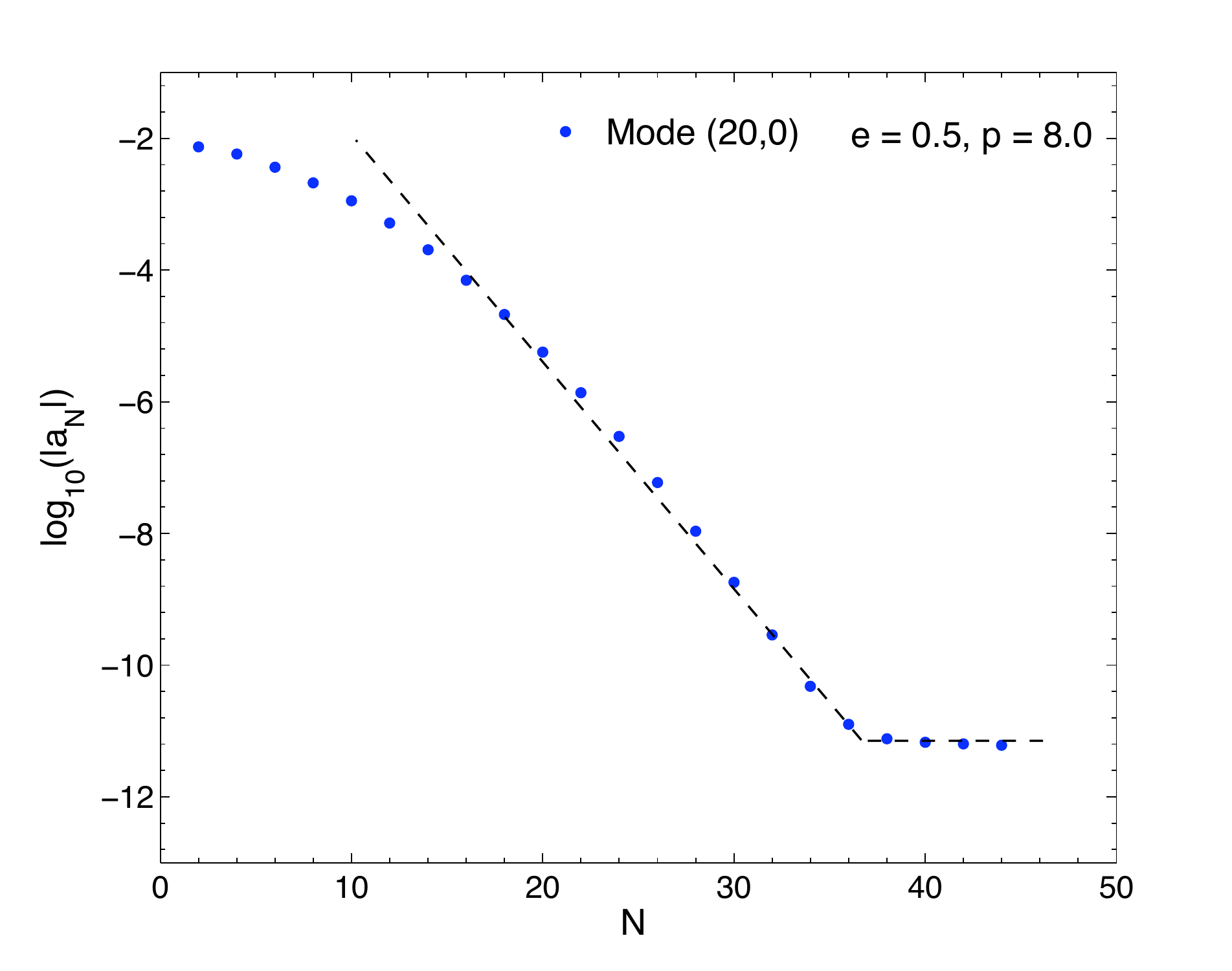}
\caption{Truncation error estimated from the absolute value of the last spectral
coefficient, $\mb{a}^{}_N$, of the variable $\psi^{\ell m} = r \,\Phi^{\ell m}$. 
We show the results for the harmonic modes $(\ell,m) = (2,2)$ (left column) and 
$(\ell,m) = (20,0)$ (right column) and for the orbital parameters $(e,p)$ $=$ $(0.1,7.0)$ 
(top row) and $(e,p)$ $=$ $(0.5,8.0)$ (bottom row), where here $e$ denotes the
eccentricity and $p$ the {\em semilatus rectum}.  We observe spectral convergence
(straight line) until we reach machine roundoff error (plateau).  The data for these 
plots has been obtained from the subdomain at the right of the particle.
\label{convergencewithparticle}}
\end{figure}

The numerical method that we use to solve our equations is the PseudoSpectral
Collocation (PSC) method~\cite{Boyd}.  We discretize each subdomain by using 
a {\em Lobatto-Chebyshev} grid.  Then, in the PSC method the variables have
a {\em spectral} representation in terms of an expansion in Chebyshev 
polynomials, $\{T^{}_{n}(X)\}$ ($X \in [-1,1]$, $n=0,\ldots,N$), and a
{\em physical} representation in terms of the values of the variables at the
collocation points: 
\begin{equation}
\mb{U}^{}_{N}(t,\rsu)~ = ~\sum_{n=0}^N \mb{a}^{}_n(t)\, T^{}_n(X(t,\rsu)) ~ 
= ~ \sum_{i=0}^N \mb{U}^{}_i(t)\, {\cal C}^{}_i(X(t,\rsu))\,,
\label{representation}
\end{equation}
where $X(t,\rsu)$ is the mapping between the spectral and physical domains
(the time dependence appears only in the eccentric case and is the way in which
we keep the particle at the interface between two subdomains), 
$\mb{a}^{}_n$ are the spectral coefficients, and ${\cal C}^{}_i$ are
cardinal functions associated to the {\em Lobatto-Chebyshev} grid~\cite{Boyd}.  
One can change from one representation to the other by means either of matrices
or fast Fourier transforms (as we do in our implementation).  An important feature 
of the PSC method is that it provides exponential convergence for smooth 
functions, which is the case of our solutions after applying the PwP formulation. 
This is illustrated in Figure~\ref{convergencewithparticle}, where we show some
convergence plots.

\begin{figure*}
\centering 
\includegraphics[width=1.0\textwidth]{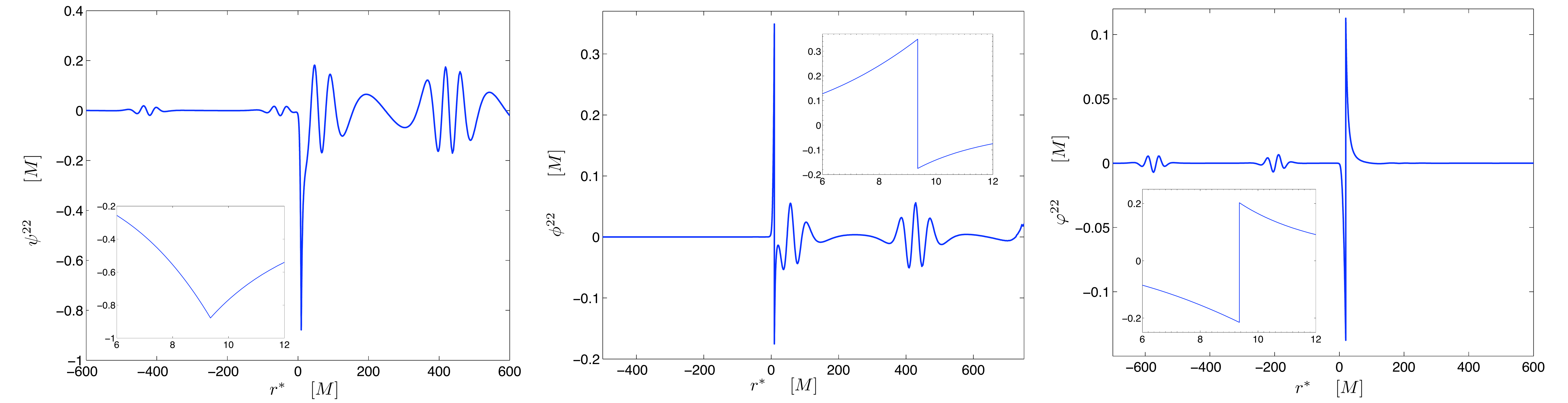}
\caption{Snapshots of the evolution of the variables $\psi^{\ell m}$ (left), 
$\phi^{\ell m}=\partial_t\psi^{\ell m}$ (center), and $\varphi^{\ell m}
=\partial_{\rsu}\psi^{\ell m}$ (right), for the mode $\ell=m=2$.  The
orbital parameters are $(e,p)$ $=$ $(0.5,8)$.   
\label{scalar_variables}}
\end{figure*}

Once we have performed the spatial discretization using the PSC method, we obtain a 
set of ordinary differential equations that can be evolved using the Method of 
Lines.  The particular implementation that we use involves a Runge-Kutta 4 solver.
The evolution must include the communication between subdomains that we have
discussed above.   We use two different numerical techniques to communicate
the subdomains:  (i) The \emph{penalty method}, where the system is driven dynamically
to satify the junction conditions, and (ii) the {\em direct communication of 
characteristic fields}, where the junction conditions are imposed by communicating
the characteristic fields of the first-order system of partial differential equations.
To illustrate the ability of our method to resolve the field and its derivatives
around the particle we show some snapshots of the evolution in Figure~\ref{scalar_variables}.

\section{Some Results and Discussion}
\label{results}

Until now we have introduced the foundations of our method and the particular numerical techniques
that we use in order to implement it. We have shown that the PwP formulation lead to smooth 
solutions at each computational subdomain, which in turn exploits the exponential convergence
of the PSC method as shown in Figure~\ref{convergencewithparticle}.  We have also shown how 
well this method can resolve the field and its derivatives around the particle location (see
Figure~\ref{scalar_variables}).  In particular, we can see how the jumps in the time and radial 
derivatives of the scalar field modes $\Phi^{\ell m}$ are well resolved (see the inset plots 
in Figure~\ref{scalar_variables} where we zoom in the area near the particle).

\begin{figure*}
\includegraphics[width=0.85\textwidth]{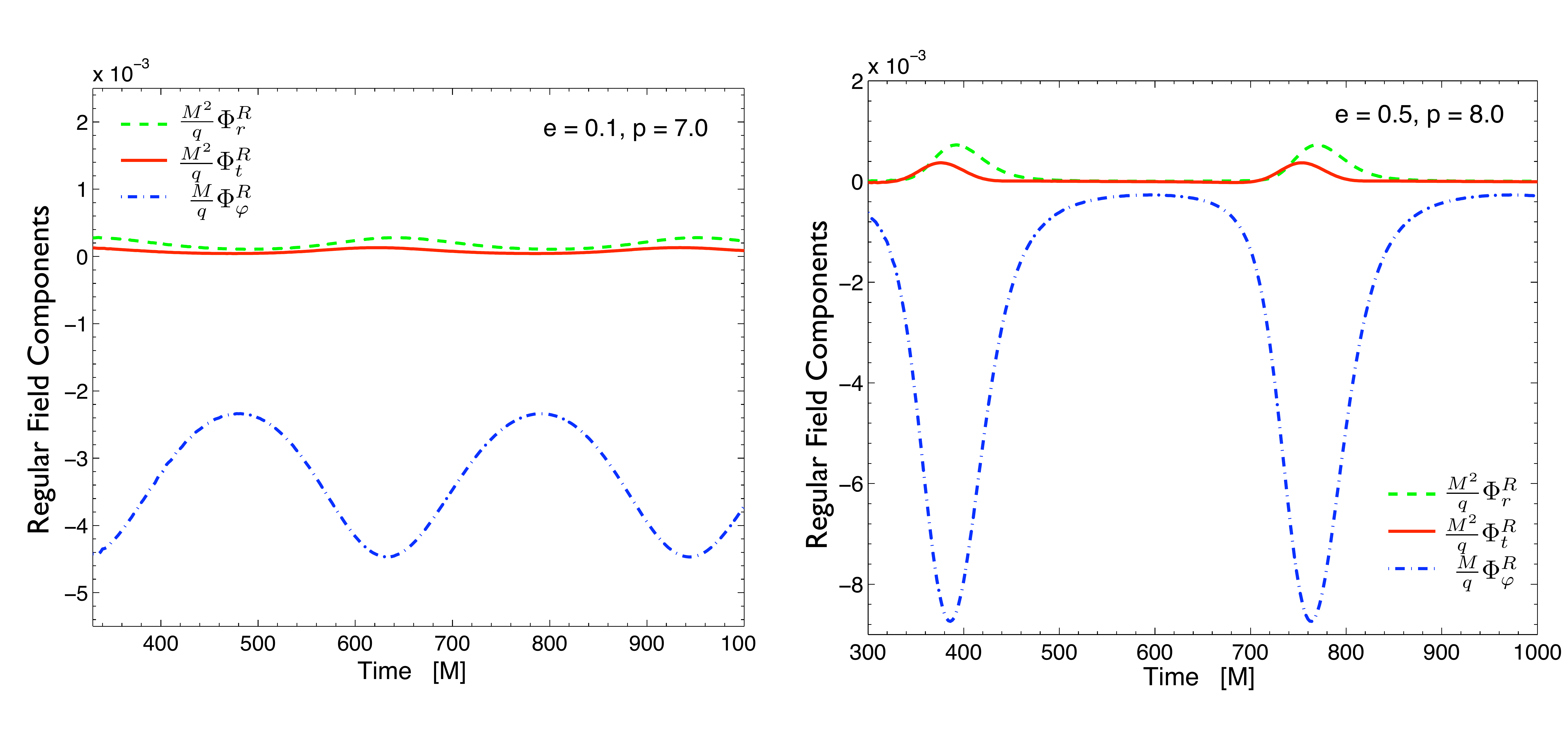}
\caption{Evolution of the components of the gradient of the regularized field, $\nabla^{}_{\alpha}\Phi^{\regu}$,
for a scalar charged particle in eccentric orbits around a non-rotating MBH.  From left to right, the orbital
parameters of the orbits are: (i) $(e,p)$ $=$ $(0.1,7.0)$ and (ii) $(e,p)$ $=$ $(0.5,8.0)$. For each orbit (frame), 
the solid line represents the evolution of the dimensionless time component, 
$\frac{M^2_{\bullet}}{q}\,\Phi^{\regu}_t$; the dashed line represents the evolution
of the dimensionless radial component, $\frac{M^2_{\bullet}}{q}\,\Phi^{\regu}_r$; and 
the dot-dashed line represents the evolution
of the dimensionless azimuthal component, $\frac{M^{}_{\bullet}}{q}\,\Phi^{\regu}_\varphi$.
\label{self_force_evolution}}
\end{figure*}

Finally, we have computed the values of the gradient of the regularized components of the gradient
of the scalar field, $\nabla_{\alpha}\Phi^R$, which contain information equivalent to the self-force 
(see Eq.~(\ref{self_force})), which is the projection orthogonal to the particle four-velocity
of $\nabla_{\alpha}\Phi^R$.  In Figure~\ref{self_force_evolution} we show the evolution of the components 
of the regularized field obtained by truncating the sum over harmonic modes in such a way that we only
include modes with $\ell\leq\ell_{max} = 20$.  This figure shows the evolution of the regularized
field for two different orbits, which 
in terms of the eccentricity and semilatus rectum are: (i) $(e,p)=(0.1,7.0)$ and (ii) $(e,p)=(0.5,8.0)$.
In order to quote some numerical results,  we have computed the values of the self-force components using
the method of the comunication of the characteristic field described in Sec.~\ref{pwpformulation}.
We obtain: $\Phi^R_t = 2.2824\times10^{-4}\,(q/M^2_{\bullet})$, 
$\Phi^R_r=  8.5836\times10^{-5}\, (q/M^2_{\bullet})$ and  
$\Phi^R_{\varphi} =  -3.7129\times10^{-3}\, (q/M^{}_{\bullet})$ 
for the eccentric orbit (i) and $\Phi^R_t =  6.4751\times10^{-6} \, (q/M^2_{\bullet})$, 
$\Phi^R_r= -5.3519\times10^{-6}\, (q/M^2_{\bullet})$ and  
$\Phi^R_{\varphi} = 2.8778\times10^{-4} \, (q/M^{}_{\bullet})$ for the eccentric orbit (ii).

All our calculations have used between 8 and 39 subdomains and 50 collocation points per domain.
The average time for a full self-force calculation (which involves the calculation of 231 harmonic
modes) in a computer with two Quad-Core Intel Xeon processors at 2.8 GHz is always in the range 
20-30 minutes.  These calculations can be further optimized by distributing the subdomains and collocation 
points so that the resolution is adapted further to our physical problem, and is this is the subject of
ongoing work.  The calculations can be easily parallelized, either by spreading the computational calculations
for individual modes or for individual subdomains.  In the last case the parallelization is not trivial as
we need to pass the relevant information for subdomain communication.

Looking to the future, we are currently working on extending these techniques to the gravitational case, 
where the challenge comes from the fact that each harmonic mode is described by a set of coupled 1+1 wave 
type equations, but where the PwP formulation can be used as it has been described here.

\section*{Acknowledgments}
PCM is supported by a predoctoral FPU fellowship of the Spanish Ministry of
Science and Innovation (MICINN).
CFS acknowledges support from the Ram\'on y Cajal Programme of the
Ministry of Education and Science of Spain and by a Marie Curie
International Reintegration Grant (MIRG-CT-2007-205005/PHY) within the
7th European Community Framework Programme. 
Financial support from the contracts ESP2007-61712 (MEC) and FIS2008-06078-C03-01 (Ministry of Science 
and Innovation of Spain) is gratefully acknowledged.
The authors have used the resources of the Centre de Supercomputaci\'o de Catalunya (CESCA)
and the Centro de Supercomputaci\'on de Galicia (CESGA; project number ICTS-2009-40).

\section*{References}

\providecommand{\newblock}{}

\end{document}